%
%

\input harvmac.tex


\def\cn{{\cal N}}

\def\IR{\relax{\rm I\kern-.18em R}}
\def\IZ{\relax\ifmmode\hbox{Z\kern-.4em Z}\else{Z\kern-.4em Z}\fi}
\def\cD{{\cal D}}

\def\tq{{\tilde q}}
\def\tQ{{\tilde Q}}

%
%
\def\np#1#2#3{{\it Nucl. Phys.} {\bf B#1} (#2) #3}
\def\pl#1#2#3{{\it Phys. Lett.} {\bf #1B} (#2) #3}
\def\plb#1#2#3{{\it Phys. Lett.} {\bf #1B} (#2) #3}
\def\prl#1#2#3{{\it Phys. Rev. Lett.} {\bf #1} (#2) #3}

\def\prd#1#2#3{{\it Phys. Rev.} {\bf D#1} (#2) #3}
\def\atmp#1#2#3{{\it Adv. Theor. Math. Phys.} {\bf #1} (#2) #3}
\def\jhep#1#2#3{{\it J. High Energy Phys.} {\bf #1} (#2) #3}
%
%


\lref\natistr{N. Seiberg, ``New Theories in Six Dimensions and 
Matrix Description of M-Theory on $T^5$ and $T^5/Z_2$,'' hep-th/9705221,
\pl{408}{1997}{98}.}

\lref\brs{M. Berkooz, M. Rozali and N. Seiberg, ``Matrix Description of 
M Theory on $T^4$ and $T^5$,'' hep-th/9704089, \pl{408}{1997}{105}.}

\lref\oens{O. Aharony, M. Berkooz, S. Kachru, N. Seiberg, E. Silverstein,
``Matrix Description of Interacting Theories in Six Dimensions,''
hep-th/9707079, \atmp{1}{1998}{148}.}

\lref\edcomments{E. Witten, ``Some Comments on String Dynamics,''
hep-th/9507121, {\it Strings '95} (World Scientific, 1996),
ed. I. Bars et. al., 501.}

\lref\edhiggs{E. Witten, ``On the Conformal Field Theory of the Higgs 
Branch,'' hep-th/9707093, \jhep{9707}{1997}{003}.}

\lref\abs{O. Aharony, M. Berkooz and N. Seiberg, ``Light Cone
Description of $(2,0)$ Superconformal Theories in Six Dimensions,''
hep-th/9712117,
\atmp{2}{1998}{119}.}

\lref\diaco{D.-E. Diaconescu and N. Seiberg, ``The Coulomb Branch of 
$(4,4)$ Supersymmetric Field Theories in Two Dimensions,''
hep-th/9707158, \jhep{9707}{1997}{001}.}

\lref\abks{O. Aharony, M. Berkooz, D. Kutasov and N. Seiberg,
``Linear Dilatons, NS Five-Branes and Holography,''
hep-th/9808149, \jhep{9810}{1998}{004}.}

\lref\sethi{S. Sethi, ``The Matrix Formulation of type IIB
Five-Branes,'' hep-th/9710005, \np{523}{1998}{158}.}

\lref\ganset{O. J. Ganor and S. Sethi, ``New Perspectives on
Yang-Mills Theories with Sixteen Supersymmetries,'' hep-th/9712071,
\jhep{9801}{1998}{007}.}

\lref\andyjuan{J.M. Maldacena and A. Strominger, ``Semiclassical Decay 
of Near Extremal Five-Branes'', hep-th/9710014, \jhep{9712}{1997}{008}.}

\lref\ahaban{O. Aharony and T. Banks, ``Note on the Quantum Mechanics
of M Theory,'' hep-th/9812237, \jhep{9903}{1999}{016}.}

\lref\dougmoore{M. R. Douglas and G. Moore, ``D-branes, Quivers and
ALE Instantons,'' hep-th/9603167.}%

\lref\berdoug{M. Berkooz and M. Douglas, ``Five-branes in M(atrix) 
Theory,'' hep-th/9610236,
\pl{395}{1997}{196}.}%

\lref\berver{M. Berkooz and H. Verlinde, ``Matrix Theory, AdS/CFT and
Higgs-Coulomb Equivalence,'' hep-th/9907100.}

\lref\seiwit{N. Seiberg and E. Witten, ``The D1/D5 System and Singular
CFT,'' hep-th/9903224, \jhep{9904}{1999}{017}.}

\lref\kpr{C. Kounnas, M. Porrati and B. Rostand, ``On $\cn=4$ Extended
Super-Liouville Theory,'' \pl{258}{1991}{61}.}

\lref\inst{N. Dorey, T. J. Hollowood, V. V. Khoze, M. P. Mattis and
S. Vandoren, ``Multi-Instanton Calculus and the AdS/CFT Correspondence
in ${\cal N}=4$ Superconformal Field Theory,'' hep-th/9901128,
\np{552}{1999}{88}.}

\lref\coleman{S. Coleman, ``$1/N$,'' in the proceedings of the 1979
Erice School on Subnuclear Physics and in ``Aspects of Symmetry,''
Cambridge University Press.} 

\lref\dps{M. R. Douglas, J. Polchinski and A. Strominger, ``Probing
Five Dimensional Black Holes with D-branes,'' hep-th/9703031,
\jhep{9712}{1997}{003}.}

\lref\zamo{A. B. Zamolodchikov, ``Irreversibility of the Flux of the
Renormalization Group in a 2d Field Theory,'' {\it JETP Lett.} {\bf 43}
(1986) 730.}

\lref\polc{J. Polchinski, ``Scale and Conformal Invariance in Quantum
Field Theory,'' \np{303}{1988}{226}.}

\lref\juan{J. M. Maldacena, ``The Large $N$ Limit of Superconformal Field
Theories and Supergravity,'' hep-th/9711200, \atmp{2}{1998}{231}.}%

\lref\gkp{S. S. Gubser, I. R. Klebanov and A. M. Polyakov, 
``Gauge Theory Correlators
from Non-critical String Theory,'' hep-th/9802109,
\plb{428}{1998}{105}.}

\lref\wittenone{E. Witten, ``Anti-de-Sitter Space and Holography,''
hep-th/9802150, \atmp{2}{1998}{253}.}

\lref\magoo{O. Aharony, S. S. Gubser, J. M. Maldacena, H. Ooguri and
Y. Oz, ``Large $N$ Field Theories, String Theory and Gravity,''
hep-th/9905111, to appear in {\it Phys. Rep}.}

\lref\bfss{T. Banks, W. Fischler, S. H. Shenker and L. Susskind,
``M Theory as a Matrix Model: a Conjecture,'' hep-th/9610043,
\prd{55}{1997}{5112}.}

\lref\motl{L. Motl, ``Proposals on Nonperturbative Superstring
Interactions,'' hep-th/9701025.}

\lref\bansei{T. Banks and N. Seiberg, ``Strings from Matrices,''
hep-th/9702187, \np{497}{1997}{41}.}

\lref\dvv{R. Dijkgraaf, E. Verlinde and H. Verlinde, ``Matrix
String Theory,'' hep-th/9703030, \np{500}{1997}{43}.}

\lref\diagom{D. Diaconescu and J. Gomis, ``Neveu-Schwarz Five-branes
at Orbifold Singularities and Holography,'' hep-th/9810132,
\np{548}{1999}{258}.}

\lref\kll{D. Kutasov, F. Larsen and R. G. Leigh, ``String Theory
in Magnetic Monopole Backgrounds,'' hep-th/9812027,
\np{550}{1999}{183}.}

\lref\gkpelc{A. Giveon, D. Kutasov and O. Pelc, ``Holography for
Noncritical Superstrings,'' hep-th/9907178.}

\lref\gk{M. Gremm and A. Kapustin, ``Heterotic Little String
Theories and Holography,'' hep-th/9907210.}

\lref\imsy{N. Itzhaki, J. M. Maldacena, J. Sonnenschein and S.
Yankielowicz, ``Supergravity and the Large $N$ Limit of Theories
with Sixteen Supercharges,'' hep-th/9802042, \prd{58}{1998}{046004}.}

\lref\bst{H. J. Boonstra, K. Skenderis and P. K. Townsend,
``The Domain Wall / QFT Correspondence,'' hep-th/9807137,
\jhep{9901}{1999}{003}.}

\lref\matder{N. Seiberg, ``Why is the Matrix Model Correct ?''
hep-th/9710009, \prl{79}{1997}{3577}.}

\lref\newseiwit{N. Seiberg and E. Witten, ``String Theory and
Noncommutative Geometry,'' hep-th/9908142.}

\lref\akisanny{A. Hashimoto and N. Itzhaki, ``Noncommutative
Yang-Mills and the AdS/CFT Correspondence,'' hep-th/9907166.}

\lref\juanrusso{J. M. Maldacena and J. G. Russo, ``Large $N$ Limit of
Noncommutative Gauge Theories,'' hep-th/9908134.}

\lref\brodie{J. Brodie, ``Two Dimensional Mirror Symmetry from M
Theory,'' hep-th/9709228, \np{517}{1998}{36}.}

\lref\alishahiha{M. Alishahiha, ``$\cn=(4,4)$ 2D Supersymmetric Gauge
Theory and Brane Configuration,'' hep-th/9710020, \pl{420}{1998}{51};
M. Alishahiha, ``On the Brane Configuration of $\cn=(4,4)$ 2D
Supersymmetric Gauge Theories,'' hep-th/9802151, \np{528}{1998}{171}.}

\lref\kapstr{A. Kapustin and M. J. Strassler, ``On Mirror Symmetry
in Three Dimensional Abelian Gauge Theories,'' hep-th/9902033,
\jhep{9904}{1999}{021}.}

\lref\natipri{N. Seiberg, private communication.}

\lref\brs{M. Berkooz, M. Rozali and N. Seiberg, ``Matrix Description of
M Theory on $T^4$ and $T^5$,'' hep-th/9704089, \pl{408}{1997}{105}.}

\lref\schoutens{K. Schoutens, ``$O(N)$ Extended Superconformal Field
Theory in Superspace,'' \np{295}{1988}{634}.}

\lref\stv{A. Sevrin, W. Troost and A. Van Proeyen, ``Superconformal
Algebras in Two Dimensional with $N=4$,'' \pl{208}{1988}{447}.}

\lref\ikl{E. A. Ivanov, S. O. Krivonos and V. M. Leviant, ``A New
Class of Superconformal Sigma Models with the Wess-Zumino Action,''
\np{304}{1988}{601}; ``Quantum $N=3$, $N=4$ Superconformal WZW
Sigma Models,'' \pl{215}{1988}{689}.}

\lref\korp{C. Kounnas, M. Porrati and B. Rostand, ``On $N=4$
Extended Super-Liouville Theory,'' \pl{258}{1991}{61}.}

\lref\chs{C. Callan, J. Harvey and A. Strominger, ``Supersymmetric
String Solitons,'' hep-th/9112030, in Trieste 1991, Proceedings, 
String Theory and Quantum Gravity 1991, 208.}

\lref\oogvaf{H. Ooguri and C. Vafa, ``Two Dimensional Black Hole and 
Singularities of CY Manifolds,'' hep-th/9511164, \np{463}{1996}{55}.}

\lref\ghm{R. Gregory, J. A. Harvey and G. Moore, ``Unwinding Strings
and T-duality of Kaluza-Klein and H-Monopoles,'' hep-th/9708086,
\atmp{1}{1998}{283}.}

\lref\paul{P. S. Aspinwall, ``Enhanced Gauge Symmetries and K3
Surfaces,'' hep-th/9507012, \pl{357}{1995}{329}.}

\lref\cghs{C. G. Callan, S. B. Giddings, J. A. Harvey and A. Strominger,
``Evanescent Black Holes,'' hep-th/9111056, \prd{45}{1992}{1005}.}

\lref\olddvv{R. Dijkgraaf, E. Verlinde and H. Verlinde, ``BPS
Quantization of the Fivebrane,'' hep-th/9604055, \np{486}{1997}{77};
``5D Black Holes and Matrix Strings,'' hep-th/9704018, \np{506}{1997}{121}.}

\lref\peetmar{D. Marolf and A. Peet, ``Brane Baldness vs. Superselection
Sectors,'' hep-th/9903213.}

\lref\michanoncom{M. Berkooz, ``Nonlocal Field Theories and the 
Non-commutative Torus,'' hep-th/9802069, \pl{430}{1998}{237}.}

\lref\minwalla{S. Minwalla and N. Seiberg, ``Comments on the IIA 
NS Five-brane,'' hep-th/9904142, \jhep{9906}{1999}{007}.}

\lref\taylor{W. Taylor, ``D-brane Field Theory on Compact Spaces,''
hep-th/9611042, \pl{394}{1997}{283}.}

\lref\suss{L. Susskind, ``T-duality in M(atrix) Theory and
S-duality in Field Theory,''
hep-th/9611164.} 

\lref\grt{O. Ganor, S. Ramgoolam and W. Taylor, ``Branes, Fluxes
and Duality in M(atrix) Theory,'' hep-th/9611202, \np{492}{1997}{191}.}


%
\rightline{RUNHETC-99-31, PUPT-1886}
\Title{
\rightline{hep-th/9909101}
} 
{\vbox{\centerline{IR Dynamics of $d=2$, $\cn=(4,4)$ Gauge Theories}
\centerline{}
\centerline{and DLCQ of ``Little String Theories''}}}

\medskip

\centerline{\it
Ofer Aharony${}^1$ {\rm and} Micha Berkooz${}^{2}$}
\bigskip
\centerline{${}^1$Department of Physics and Astronomy, 
Rutgers University, Piscataway, NJ 08855}
\centerline{${}^2$Department of Physics, Jadwin Hall, Princeton University, 
NJ 08544}

\smallskip

\vglue .3cm
\bigskip

\noindent
We analyze the superconformal theories (SCFTs)
which arise in the low-energy limit of
$\cn=(4,4)$ supersymmetric gauge theories in two dimensions, primarily the
Higgs branch SCFT. 
By a direct field theory analysis we find a continuum of
``throat''-like states localized near the singularities of the Higgs
branch. The ``throat'' is similar to the ``throat''
found in the Coulomb branch of the same theories, but the full superconformal
field theories of the two branches are different.
A particular example is the SCFT of the $\IR^4/\IZ_2$ sigma model with
zero theta angle.
In the application of the Higgs branch SCFTs to the DLCQ description of
``little string theories'' (LSTs), the ``throat'' continuum is identified with
the continuum of ``throat'' states in the holographic description of
the LSTs. We also match the descriptions of the string interactions
(in the ``throat'' region) in the DLCQ and holographic descriptions of
the $\cn=(2,0)$ LSTs.

\Date{9/99}

\newsec{Introduction}

Gauge theories with eight supercharges in various dimensions have been an
interesting subject for several years now, mostly because supersymmetry
severely restricts the quantum corrections in these theories and
allows many exact computations to be made. 
The two dimensional case is especially interesting because these gauge
theories appear in linear sigma model descriptions of some string theory
compactifications and in discrete light-cone (DLCQ) descriptions of
``little string theories'' (LSTs).

The moduli space of gauge theories with eight supercharges
(below 6 dimensions) has two general types of branches : the
Coulomb branch, where the scalars in the vector multiplets obtain
vacuum expectation values, and the Higgs branch, where the scalars in the
hypermultiplets obtain vacuum expectation values.
To the extent that the metric on the moduli space is a useful notion (i.e., it
is the leading term in a systematic expansion), it is very much constrained by
supersymmetry. In theories with 8 supercharges,
supersymmetry 
forbids any quantum corrections to the Higgs branch moduli space metric. On
the Coulomb branch, in perturbation theory only one-loop corrections
to the metric are possible, and often the non-perturbative
corrections are also understood (or forbidden altogether). 
Thus, it is generally possible to compute
exactly the metric on the moduli space in these theories. 
Naively, this means that we have good control over the low-energy behavior in 
these theories. However,
there are some cases where the moduli space approximation breaks down, and
more information is needed to give a full low-energy description. Although this
break-down can happen in a variety of ways, which are generally indicated
by higher-order terms in the effective action becoming important,
it certainly occurs when
the metric becomes negative, and often also when it is singular.

In gauge theories with 8 supercharges, such singularities of the metric occur 
near the meeting points of the Higgs and Coulomb
branches, where additional degrees of freedom classically (and in some
cases also quantum mechanically) become massless. 
The significance of the singularities changes in different dimensions.
Above two dimensions,
there is a real moduli space of vacua, and
once a point on the moduli space is chosen
the theory is well-described (at low enough energies) by fluctuations on 
the moduli space around that point.
Away from singular points in the moduli space the theory 
is typically free in the IR, while at singular points it is typically described
by a non-trivial superconformal field theory (SCFT).
In two or less dimensions
the situation is different, since there is no real moduli space of vacua
due to quantum fluctuations (the moduli space metric is still often
relevant as part of a Born-Oppenheimer approximation). Singularities in
the metric
indicate that the sigma model 
approximation breaks down
(although one can sometimes still obtain sensible answers for a 
carefully chosen restricted 
class of questions). 

In this paper we will discuss the $1+1$ dimensional
superconformal field theories which are the
low-energy limit of $\cn=(4,4)$ supersymmetric gauge theories in two dimensions.
The
low-energy dynamics in this case is described by two 
superconformal field theories,
one corresponding to the Higgs branch and one to the Coulomb branch; the
two theories decouple at low energies \refs{\edcomments,\oens,\edhiggs}. 
The moduli space metrics of the two
IR theories have singularities (and can even become negative on the Coulomb
branch), indicating that higher
order terms are important, and one requires a different description of the
theory near the singular regions.
In particular, there are various
reasons to believe that there is a continuous spectrum associated with
the singularities in the Higgs branch (one example of such a singularity
is the $\IR^4/\IZ_2$ singularity with zero theta angle). In this paper we
will develop, following \refs{\edhiggs,\berver}, 
a field theory method which yields a
useful description of 
the behavior of the theory
near the singularities. This is done by means of a 
simple effective field theory 
which explicitly exhibits the continuous
spectrum, and in some cases also
the leading interaction term
in a systematic expansion. 

The method we use involves an explicit Lagrangian for the Higgs branch
SCFT, in which the Coulomb branch fields appear as auxiliary variables. We
can then integrate out the hypermultiplets to obtain an effective action
for these auxiliary fields.
This method was outlined in \berver. The main 
difference will be that whereas there the 
Hamiltonian formulation was emphasized 
(which is more convenient in the case of quantum 
mechanics), here we will focus on the Lagrangian approach, and we will 
take advantage of the power of $1+1$ dimensional superconformal symmetry. 
The approach we use is
also similar to that of \inst, although the interpretation is quite different,
and the methods we use are similar to those of \kapstr.

In our description 
the singularity will be replaced by a semi-infinite ``throat''\foot{
For the case of $\IR^4/\IZ_k$ singularities embedded in a space with a 
compact
circle transverse to the singularity, 
there is a T-duality relating them to an NS5-brane configuration
which has a similar semi-infinite ``throat'' \refs{\oogvaf,\ghm}.},
as
in the Coulomb branch when the
one-loop corrections are taken into account \diaco. A similar ``throat''
was found \seiwit\ in the analysis of D1-D5 theories which have the same
singularities as the Higgs branch theories we discuss here;
however,
their analysis used the AdS/CFT correspondence and a large $N$
approximation, while our analysis will be directly in the field theory and
valid for any value of $N$. Furthermore, our method is applicable for any 
Higgs branch of a $\cn=(4,4)$ theory (although we will focus on $U(N)$
gauge theories, which are related to the
D1-D5 system), and we will discuss both the R-R and the NS-NS sectors of
the theory.

We will focus mainly on the $U(N)$ gauge theory with an adjoint
hypermultiplet and $N_f$ fundamental hypermultiplets (which
arises as the low-energy theory on D1-branes inside D5-branes in flat space).
One of our motivations for studying this theory is
to examine its usefulness as a DLCQ description \refs{\oens,\edhiggs}
of ``little string theories''
(LSTs) with 16 supercharges \refs{\natistr,\brs,\olddvv}.
The ``little string theories'' are known from their holographic description
(which includes a linear dilaton background)
to have a continuous spectrum above some mass gap \refs{\andyjuan,\abks}. 
We will show that in
the DLCQ this continuous spectrum may be identified with the continuous spectrum
appearing at the singularities in the corresponding SCFTs, and that we can
also reproduce the string interactions in the weakly coupled region of
the holographic description. 

The organization of the paper is as follows. In section 2 we review 
$\cn=(4,4)$ gauge theories in two dimensions and
their low-energy limits\foot{
These theories were analyzed using brane constructions in \refs{\brodie,
\alishahiha}, and were argued to be dual to the dimensional
reduction of some non-trivial $2+1$ dimensional field theories in \sethi.},
and we give an explicit Lagrangian description
for the conformal theory of the Higgs branch which is also useful near the
singularities. In section 3 we analyze the
behavior of this theory near the singularities of the Higgs branch in
$U(N)$ gauge theories, and
show that it develops a ``throat'' region which can be explicitly
exhibited and analyzed.
In section 4
we analyze the DLCQ of the $(2,0)$ LSTs, which corresponds to a 
particular Higgs
branch SCFT. We match the continuous part of the spectrum to that 
which arises in the linear dilaton region of the holographic description
of the LSTs, 
and we also reproduce the string interactions there. 
In
section 5 we analyze the DLCQ of $(1,1)$ LSTs \refs{\sethi,\ganset}, 
which is a particular
Coulomb branch SCFT. In this case we can easily exhibit the low-energy states
of the $(1,1)$ LSTs (which are free W-bosons), but it is more difficult to
obtain a precise understanding of the continuum states since the ``throats''
of the relevant
Coulomb branch theories are less well understood. We end in section 6
with a summary and some remarks.

\newsec{Two Dimensional $\cn=(4,4)$ Gauge Theories and their 
Low Energy Limits}

\subsec{$\cn=(4,4)$ Gauge Theories}

Gauge theories with $\cn=(4,4)$ supersymmetry in two dimensions may be
viewed as the dimensional reduction of four dimensional $\cn=2$ gauge
theories, or of six dimensional $\cn=(1,0)$ gauge theories. Like the
other theories with eight supercharges, their matter content includes
vector multiplets in the adjoint representation of the gauge group and
hypermultiplets which can be in arbitrary representations of the gauge
group. The six dimensional gauge theories have an $SU(2)_R$
R-symmetry. Upon dimensional reduction to two dimensions there is an
additional $SO(4)
\simeq SU(2)_r \times SU(2)_l$ symmetry acting on the four reduced
dimensions. This is also an R-symmetry since the supercharges are a
spinor of this $SO(4)$ group; the left-moving (positive chirality)
supercharges are in the $\bf{(2,1,2)}$ representation of
$SU(2)_l\times SU(2)_r \times SU(2)_R$ while the right-moving
(negative chirality) supercharges are in the $\bf{(1,2,2)}$
representation. 

The matter content of the vector multiplets and the
hypermultiplets may be easily
derived using the dimensional reduction from six dimensions :
\eqn\mltplts{\matrix{
      &          & SU(2)_l & SU(2)_r & SU(2)_R &\cr
vector\ multiplet: & A_\mu    &       1 &       1 & 1       & vector\cr
      & V        &       2 &       2 & 1       & real\ scalar\cr
      & \psi^L_V &       1 &       2 & 2       & left\ moving\ fermion \cr
      & \psi^R_V &       2 &       1 & 2       & right\ moving\ fermion \cr
hypermultiplet: & H        &       1 &       1 & 2       & complex\ scalar\cr
      & \psi^L_H &       2 &       1 & 1       & complex\ left\ moving\ fermion\cr
      & \psi^R_H &       1 &       2 & 1       & complex\ right\ moving\ fermion.\cr }}

The parameters of $\cn=(4,4)$ gauge theories in two dimensions include
the gauge coupling (or gauge couplings for non-simple gauge groups),
Fayet-Iliopoulos terms and theta angles
for Abelian components of the gauge group, and
masses for the hypermultiplets.  We will discuss here mostly the case
where the Fayet-Iliopoulos terms, theta angles and masses 
all vanish, so that the
only parameter is the gauge coupling constant $g_{YM}$, whose
scaling dimension is $[g_{YM}] = 1$.  Because of this scaling
dimension, the gauge theories become free in the UV, but are
strongly coupled in the IR, where the perturbative gauge theory
description is not valid.
Schematically, the gauge theory
Lagrangian (suppressing all indices) is given by
\eqn\gaugetheory{\eqalign{
{\cal L} = \int d^2x \{ & {1\over {4g_{YM}^2}} \tr \left[ F_{\mu \nu}^2 +
({\cal D}_\mu V)^2 + [V, V]^2 + \bar{\psi}_V \gamma^\mu
\cD_\mu \psi_V + \bar{\psi}_V [V, \psi_V] + D^2 \right]
+ \cr
& \sum_{hypers} \left[ 
(\cD_\mu H)^2 + (VH)^2 + \bar{\psi}_H \gamma^\mu \cD_\mu \psi_H +
\bar{\psi}_H V \psi_H + \bar{\psi}_V H \psi_H + D H H \right] \}, }}
where $D$ are the three auxiliary fields in the vector multiplet (in
the $\bf 3$ of $SU(2)_R$) and
${\cal D}_\mu \simeq \del_\mu + A_\mu$ is the covariant derivative.
The scaling dimensions of the various fields in this Lagrangian are
$[A_\mu] = [V] = 1$, $[\psi_V] = 3/2$, $[D] = 2$, $[H] = 0$ and
$[\psi_H] = 1/2$.

The classical moduli space of these theories, like that of other
theories with eight supercharges (below six dimensions), includes two
types of configurations : the Coulomb branch, where the scalars in the
vector multiplets obtain vacuum expectation values (VEVs), and the Higgs branch,
where the scalars in the hypermultiplets obtain VEVs (in
general there are also mixed branches but they do not seem to raise
any new issues so we will limit ourselves here to a discussion of the two
extreme cases). Of course, quantum field theories in two dimensions do not
actually have a moduli space of vacua, since wave functions tend to
spread out on the classical moduli space; hence, one needs to treat the
IR limit more carefully.

\subsec{Basic Properties of the IR Limit(s)}

The low-energy (IR) limit of the gauge theory involves taking $g_{YM}
\to \infty$, and one expects to obtain in this limit 
an $\cn=(4,4)$ superconformal 
field theory (SCFT)\foot{The IR theory 
is (by definition) a scale-invariant theory. 
For theories with a compact moduli space this
implies \refs{\zamo,\polc} 
that it is also a conformally invariant theory; we
expect this to be true also in our case, even though the moduli space
is non-compact.}.
However,
it is believed
\refs{\edcomments,\dps,\oens,\edhiggs} that the low-energy limit is in fact
two decoupled superconformal field theories, one describing the Higgs
branch of the theory and the other corresponding to the Coulomb
branch; for finite $g_{YM}$ wave functions can spread from one branch
to the other, but it is believed that in the $g_{YM} \to
\infty$ limit the distance between the branches goes to infinity and
they decouple. 

The simplest general argument for this \edhiggs\ comes from an
analysis of the R-symmetries of the IR $\cn=(4,4)$ SCFT. The
$\cn=(4,4)$ superconformal algebra includes left-moving and
right-moving $SU(2)$ Kac-Moody algebras. Both the Higgs and the
Coulomb branches have semi-classical regions for large VEVs, whose
description includes scalar fields with a flat metric; a symmetry
which rotates such scalar fields cannot be separated into left-moving
and right-moving pieces, so it cannot be part of the R-symmetry of the
IR SCFT. This determines that in a SCFT of the Higgs branch
$SU(2)_R$ cannot be (part of) an R-symmetry, since the scalars in the
hypermultiplets are charged under this group. Similarly, $SU(2)_r\times
SU(2)_l$ cannot be an R-symmetry in the theory of the Coulomb
branch. It is natural to expect that the R-symmetry of the Higgs
branch is exactly $SU(2)_r\times SU(2)_l$, while the R-symmetry of the
Coulomb branch is an $SU(2)_1\times SU(2)_2$ symmetry which is not
visible in the gauge theory but which has $SU(2)_R$ as its diagonal
subgroup. Since the Coulomb and Higgs branch superconformal theories
have different R-symmetries it is clear that they cannot be
identified. In the Higgs branch SCFT we can use the identification of
the $SU(2)$ R-symmetries and the relation between their chiral anomaly
and the central charge to compute the central charge, $c = 6(n_H - n_V)$
where $n_H$ is the number of hypermultiplets and $n_V$ is the number of
vector multiplets \edhiggs.

As discussed in \edhiggs, the Higgs and Coulomb branch theories have
different scaling
dimensions for the various fields. In the Higgs branch the
hypermultiplets obtain VEVs so $[H] = 0$, while in the Coulomb branch
the vector multiplet scalars obtain VEVs so $[V] = 0$. 
In the Higgs branch theory, the scaling dimensions of the various
fields agree with their classical scaling dimensions in
\gaugetheory; the fact that $[V]=1$ follows from the fact that it is in
the $\bf(2,2)$ representation of the $SU(2)\times SU(2)$ 
R-symmetry group in the $\cn=(4,4)$
superconformal algebra.
Thus, we claim (following \refs{\berver,\edhiggs}) 
that the exact Lagrangian for the
Higgs branch SCFT is obtained
by the naive limit of $g_{YM} \to \infty$ in \gaugetheory, which
removes the first line and leaves us only with the second line :
\eqn\higgs{
{\cal L}_H = \int d^2x \sum_{hypers} \{ 
(\cD_\mu H)^2 + (VH)^2 + \bar{\psi}_H \gamma^\mu \cD_\mu \psi_H +
\bar{\psi}_H V \psi_H + \bar{\psi}_V H \psi_H + D H H \}. }
In
this limit the kinetic terms of the vector multiplet fields all
vanish, so they become auxiliary fields. 

One description of the Higgs branch SCFT is as a supersymmetric sigma model
on the Higgs branch moduli space : this is recovered if
one integrates out the auxiliary fields in \higgs. This can be done explicitly 
since the action is quadratic in these fields. The
integration over the $D$ fields and the gauge fields gives the
constraints which limit the configuration space to the classical Higgs
branch, while the integration over $V$ gives rise to the
4-fermi interactions of the supersymmetric non-linear sigma model. 
However, following \berver, in order to describe the theory near the
singularities in the moduli space
it is more useful to regard the vector
multiplet fields, which (using \higgs)
are composite operators on the Higgs branch of the form $V \sim 
\psi_H^L \psi_H^R / H^2$, 
as the basic objects, and to perform 
a change of variables to these new coordinates. Note that these variables
are all invariant under any global symmetries acting on the hypermultiplets,
so we cannot describe states which are charged under such global symmetries
in terms of the new variables.

In the Lagrangian approach this procedure amounts to integrating over 
the hypermultiplet fields
and inducing an effective action for the vector multiplet fields \edhiggs.
This is basically the same technique as in the ${\bf CP}^{N-1}$ model 
(described in \coleman\ and
references therein). Obviously, in this approach we are throwing away
most of the dynamical degrees of freedom of the theory, and naively
one might expect the resulting (non-local)
effective action to be uncontrollable.
However, 
since a non-zero value of $V$ gives a mass to $H$ and $\psi_H$,
there is actually a systematic 
expansion in $E/V$ (where $E$ is the energy), 
which gives a good description of the states which
are localized near the singularities. 
This will be carried out in section 3, and used for
DLCQ applications in section 4.

When is this description expected to be valid ? 
Using \higgs\ we can
think of $V^2$ as the mass squared of the hypermultiplets, so we
expect such a description to be valid for energies below this mass
scale. This is just the usual Born-Oppenheimer approximation.
Alternatively, the effective action for the vector multiplets
may be expanded in a power series in the dimensionless
$dV/V^2$, and we can neglect
the higher order terms in this expansion when $dV \ll V^2$. Thus, we
expect to get a good description in terms of the vector multiplets for
low-energy wave functions concentrated around large values of $V$. If we
integrate out the vector multiplets instead of the hypermultiplets, we
find that $V$ is equal to a bilinear in the hypermultiplet fermions
divided by $H^2$, so in some sense large values of $V$ correspond to
small values of $H$, close to the singularity of the Higgs
branch\foot{Related observations were made in \peetmar.}. 
Therefore, we expect the effective vector multiplet theory to give a good
description of the region near the singularities of the Higgs
branch. 

If, alternatively, we want
to focus on the Coulomb branch theory, we would like to 
have $[V] = 0$, so we
need to take $V \to \infty$ as $g_{YM} \to \infty$ keeping the
dimensionless $\Phi \equiv V/g_{YM}$ constant. This will be the coordinate on
the moduli space of the Coulomb branch SCFT. Note that this means
that any finite value of $\Phi$ in the Coulomb branch theory
corresponds to an infinite $V$ from the point of view of the original
gauge theory and of the Higgs branch theory, which is consistent with
the infinite separation between the two branches in the IR. 
It is not clear whether the same procedure we used for analyzing 
the Higgs branch can also be used for the Coulomb branch,
i.e., whether we can treat the
hypermultiplets as 
auxiliary variables in the Coulomb branch SCFT and use them to understand
its singularities. We have not been able to write down an explicit Lagrangian
for the Coulomb branch SCFT, analogous to \higgs.
The approach we will use to analyze the Coulomb branch SCFT in 
section 5 will be to first integrate out the
hypermultiplets for finite $g_{YM}$, and then take $g_{YM} \to \infty$
with the scaling which keeps $\Phi$ finite. This scaling keeps both
the tree-level and the one-loop terms in the Coulomb branch moduli
space metric finite.

\newsec{The ``Throat'' of the Higgs Branch}

The moduli space metric on the Higgs branch does not
receive any quantum
corrections. Thus, unlike the situation in the Coulomb branch, which 
we will discuss in section 5, the singularities in the Higgs branch
(where classically it meets the Coulomb branch)
remain a finite
distance away. Nevertheless, there are many reasons to believe that
the moduli space description breaks down near the singularities
\refs{\edcomments}. A special case of such a singularity, which arises
in the $U(1)$ theory with two hypermultiplets, is the $\IR^4/\IZ_2$
singularity with zero theta angle.
When this SCFT is used for string theory compactifications, some
correlation functions diverge, as is evident from the existence of
light non-perturbative states in space-time.
In \seiwit\ it was claimed, using the AdS/CFT
correspondence \refs{\juan,\gkp,\wittenone,\magoo}, that at the singularities
the Higgs
branch develops a semi-infinite ``throat'' in appropriate
variables. Such a ``throat'' (which appears also in the perturbative analysis
of the Coulomb branch metric \diaco) leads to a continuous spectrum
of dimensions of primary operators (beyond the continuum associated with the
classical non-compactness of the moduli space) and to divergences in various
correlation functions.
In this section we will discuss a field theory analysis of the
singularities which, among other things,
reproduces this claim directly 
(without using a large gauge group approximation as
in \seiwit), as the leading term in a systematic expansion.

\subsec{The ``Throat'' in the Abelian Case}

Let us begin by analyzing the superconformal theory corresponding to
the Higgs branch of a $U(1)$ gauge theory with $N_f$ charged
hypermultiplets, with $N_f > 1$ (otherwise there is no Higgs
branch). This is an $\cn=(4,4)$ SCFT with central charge $c=6(N_f-1)$.
As explained before, we would like to obtain an effective description
of the singularity in terms of the vector multiplet fields. This
effective theory is expected to give a good description of some states
which are localized near the singularities.  We begin with the case of
a $U(1)$ gauge field as the easiest implementation of our approach,
and discuss the additional features of the non-Abelian $U(N)$ gauge
theory in the next subsection. The $U(1)$ case is also interesting in
its own right since, as mentioned above, the $N_f=2$ Higgs branch
theory gives the $\IR^4/\IZ_2$ sigma model with zero theta angle.

\medskip
\noindent{\it The Effective Theory}
\medskip

We would like to integrate out the fundamental hypermultiplets and obtain an
effective action for the vector multiplet
fields. As explained above, the appropriate
expansion is in terms of $d V / V^2$. 
The leading terms in this expansion, which are 
the metric term and its supersymmetric
partners, are known because of supersymmetry, and they are expected
to dominate at energies much smaller than $V$.
Supersymmetry and $SO(4)$ symmetry \refs{\dps,\diaco}
constrain the metric for the four scalar fields in the vector
multiplet to be of the form 
\eqn\metric{ds^2 = \left( a + {b \over { 2|V|^2}} \right) dV^2.}
Since
$[V]=1$, $a$ is dimensionful and so must be zero in the superconformal
theory we are discussing. On the other hand, $b$ can be non-zero, and a
1-loop computation using either the original action or the action
\higgs\ gives\foot{
It is clear that $b$ must be proportional to
$N_f$, since the whole induced action for the vector fields is
proportional to $N_f$, arising from integrating over $N_f$ independent
hypermultiplets. This form of the metric also follows (up to a constant)
from conformal invariance, since $[V]=1$.}
 $b = N_f$. In the original theory dimensional analysis
determines that $b$ cannot depend on $g_{YM}$ so the one-loop result
for $b$ must be exact; we claim that this is true also in the IR
theory \higgs. Similarly,
a torsion term also arises at one-loop \diaco.

Together, these terms lead to a ``throat'' metric.
Changing to radial
coordinates $dV^2 = dv^2 + v^2 d\Omega_3^2$, and defining a new variable
$\phi =
\sqrt{N_f/2} \log(v/M)$ for some mass scale $M$, we find a sigma model
with metric 
\eqn\newmtr{ds^2 = d\phi^2 + {N_f\over 2} d\Omega_3^2}
 and torsion 
\eqn\newtor{H = -N_fd\Omega,}
where $d\Omega_3^2$ is the metric and
$d\Omega$ is the volume form on the 3-sphere. The
metric and torsion on the $S^3$ coordinates give exactly (after adding
in the fermions) a level $N_f$ supersymmetric $SU(2)$ WZW model. From the
construction it is clear that the left-moving and right-moving $SU(2)$
symmetries of this model are part of the $SU(2)_r \times SU(2)_l$
symmetries (acting on $V$) 
that are in the superconformal algebra; we will discuss the
exact form of the $SU(2)$ currents below. 

However, the description above cannot be the whole
story. The simplest way to see this is to note that it does not have
the correct central charge, $c=6(N_f-1)$. The ``throat'' theory
describes the region of very large $\phi$, where it differs from the
full superconformal theory only by higher derivative terms that do not
contribute to the conformal anomaly, so it must have the same
conformal anomaly as the full theory. We would like to argue that the
central charges are matched by having in the effective Lagrangian a
background charge for $\phi$, of the form ${\cal L} \sim {Q\over 8\pi}
\phi \sqrt{g} {\cal R}$ (where ${\cal R}$ is the scalar curvature of
the background 2-dimensional metric $g$), with
\eqn\crg{Q=(N_f-1) \sqrt{2\over N_f}.}
The arguments for this are the following:
\item{1.}
In the original action \higgs,
a conformal transformation takes $V \to e^{\alpha} V$ (along with some
action on the fermions and the auxiliary fields), and
is known to change the action 
by ${c\over {48\pi}} \alpha \sqrt{g} {\cal R} = {{N_f-1} \over {8\pi}}
\alpha \sqrt{g} {\cal R}$. After
the change of variables, this transformation acts on $\phi$ as $\phi
\to \phi + \alpha
\sqrt{N_f/2}$;
thus, to get the correct variation of the action, it must
include a term of the form ${{N_f-1} \over {8\pi}} \sqrt{2\over N_f}
\phi \sqrt{g} {\cal R}$, giving \crg. This is nothing but the usual argument of
adding a background charge to obtain the correct central charge.
Using the fact that the supersymmetric level $N_f$ $SU(2)$ WZW model
may be written as the sum of a bosonic level $(N_f-2)$ $SU(2)$ WZW
model plus three free fermions, we find that in order to have the
correct central charge we require
\eqn\centcharge{6(N_f-1) = 2 + {{3(N_f-2)}\over N_f} + (1 + 3Q^2),}
giving $Q = (N_f-1) \sqrt{2\over N_f}$ (as in \diaco, the positive sign is
determined by identifying the strong coupling region with the region
near the singularity).
\item{2.}
The ``throat'' theory (ignoring all the higher order corrections) in
fact has a large $\cn=4$ algebra \refs{\schoutens,\stv,\ikl,\kpr}, 
as reviewed for example in
section 4 of \seiwit. This large $\cn=4$ algebra has two
superconformal $\cn=4$ subalgebras, one with $c=6$ and the other with
$c=6(N_f-1)$. The first algebra appears in the ``throat'' which is
found in the Coulomb branch
\diaco; in this algebra, the $SU(2)$ currents (of level 1) do not
involve the bosonic $SU(2)$ WZW model but only the free fermionic
fields (recall that these are, for the left-movers, in the
$\bf(1,2,2)$ representation of $SU(2)_l\times SU(2)_r \times SU(2)_R$,
and that $SU(2)_R$ is related to the superconformal algebra of the Coulomb
branch). However, it is the other $\cn=4$ subalgebra which is relevant
for the Higgs branch, for example because it includes the $SU(2)_r
\times SU(2)_l$ currents of level $(N_f-1)$
which are in the superconformal algebra of
the Higgs branch (in the ``throat'' theory these currents are the sum
of the bosonic level $(N_f-2)$ currents and a level one current from
the free fermions). The difference between the two subalgebras
involves exactly a shift in the energy-momentum tensor which changes
the background charge of the $\phi$ field from the value ${\tilde
Q}=-\sqrt{2/N_f}$ of the Coulomb branch ``throat'' \diaco\ to the value
$Q=(N_f-1)\sqrt{2/N_f}$ of the Higgs branch ``throat''. 

Thus, we conclude that the effective theory in the region of large $V$
is given by four free fermions, a free scalar with background charge
$Q = (N_f-1)\sqrt{2/N_f}$, and a level $(N_f-2)$ bosonic $SU(2)$ WZW model.
For $N_f=2$ this ``throat'', with its continuum of states (given by
states with any momentum in the $\phi$ direction), is part of the 
$\IR^4/\IZ_2$ sigma model with zero theta angle.

\medskip
\noindent{\it Relation to the Coulomb Branch Throat}
\medskip

The theory we found in the ``throat'' seems very similar to the theory
found in the Coulomb branch ``throat'' of the same gauge theories
\diaco, except for the
different background charge for the scalar\foot{
Corrections to the ``throat'' in the two cases will be very different,
occurring at small $V$ for the Higgs branch ``throat'' and at large $V$
for the Coulomb branch ``throat''.}. This similarity should not be
surprising, since before we take the low-energy limit the two
``throats'' are actually connected to each other, and we can have
states which start out as some wave packet (with energy well
below $g_{YM}$) in the Coulomb branch
moving towards the singularity, then enter the ``throat'' and
eventually come out of the ``throat'' into the Higgs branch. 
This suggests that there should be a one-to-one mapping between the
states propagating in the ``throat'' regions of the two theories,
although there is no direct relation between the ``throats'' after
taking the IR limit.
Studying this mapping will lead us to discover an extra vacuum energy
in the ``throat'' of the Higgs branch SCFT on a cylinder (in the RR
sector).

Let us look at the finite coupling gauge
theory on a cylinder with periodic boundary conditions for the
fermions (the RR sector). At energies well below the Yang-Mills coupling (but
still above a mass gap which will be discussed shortly,
which is a number times the inverse radius of the circle), there are four
distinct regions in the moduli space :
\item{(1)} The asymptotic flat region of the Coulomb branch.
\item{(2)} The ``throat'' of the Coulomb branch, which includes a scalar 
with background
charge $\tQ=-\sqrt{2/N_f}$.
\item{(3)} The ``throat'' of the Higgs branch, which includes a scalar 
with background
charge $Q=(N_f-1)\sqrt{2/N_f}$.
\item{(4)} The asymptotic region of the Higgs branch.

We would like to compute the energy gap between the lowest states in
regions (3) and (4), which become part of the Higgs branch SCFT when we
take the low-energy limit. In all regions there is no normalizable ground state,
but there are delta-function normalizable states whose energy is bounded
from below, and we will refer to this lower bound as the ``ground state''
energy of the appropriate region. 

We begin by reviewing the energies of states in general ``throat'' (linear
dilaton) CFTs.
In a ``throat'' region, if we normalize the state corresponding to the
identity operator to have zero energy, then
the energy of a state with
momentum $q$, corresponding to an operator $e^{iq\phi}$, is given by
$E = q(q+iQ)$ (in units determined by the radius of the cylinder). In
order to have a real energy we require $q = -iQ/2 + q_0$ where $q_0$
is real, and then $E = q_0^2 + Q^2/4$, leading to an energy gap 
(``ground state'' energy) of
$Q^2/4$ for states propagating in the ``throat'' (and a continuum
above this gap). Of course, ``throat'' states can also include some
operators involving the fermions and the level $(N_f-2)$ WZW model,
but we will ignore these for now (since these sectors are the same in
the Coulomb branch and Higgs branch ``throats''). 

We will now compute the ``ground state'' energies in regions
(1), (2) and (4), and then by matching the ``throat'' states we will compute
the ``ground state'' energy in region (3).
In regions (1) and (4) supersymmetry tells us that the ``ground state''
energies (in the RR sector) vanish. In region (2) the theory is just
a ``throat'' theory \diaco\ given in terms of the original Coulomb branch
variables (with no additional energy shifts), so the ``ground state''
energy there is
$\tQ^2/4$. We claim that this can be identified with the ``ground state''
energy in the
Higgs branch ``throat'', since states going into one come out of the
other (the scalar $\phi$ in the ``throat'' theory is free, 
and this approximation becomes
better and better ``down the throat''). Thus, we conclude that the 
Higgs branch ``throat'' has a constant contribution $\tQ^2/4-Q^2/4$ to its
vacuum energy, so that the Hamiltonian for the Higgs branch ``throat'' is
given by (ignoring the fermions and the WZW model)
\eqn\higgsham{H = q(q+iQ) + (\tQ^2-Q^2)/4.} 
Using this Hamiltonian, any state
corresponding to an operator $e^{i\tq \phi}$ and a wave-function
$e^{i(\tq+i\tQ/2)\phi}$ in the Coulomb branch ``throat'' can be
mapped into a state with the same wave-function and energy in the Higgs
branch ``throat'', corresponding to the operator $e^{iq\phi}$ where
$q+iQ/2=\tq+i\tQ/2$.

The discussion in the previous paragraph may appear to be at odds with
the discussion of \seiwit, where it was argued that the Higgs branch
``throat'' {\it does} have a mass gap of $Q^2/4$ (and not $\tQ^2/4$ as
we find above) above the ``ground state'' energy in region (1). 
However, the discussion of \seiwit\ was in the NS-NS
sector of the superconformal field theory and not in the R-R sector which we
analyzed above (in the AdS$_3$/CFT$_2$ correspondence, the NS-NS
sector vacuum is identified with the AdS$_3$ vacuum, while the R-R
vacuum is identified with the BTZ black hole). 
The vacuum (``ground state'') 
energy of the NS-NS sector is shifted compared to the R-R
sector by $(-c/12)$. Thus, if we were to repeat the same analysis we did
above in the NS-NS sector, we would find that the Coulomb branch
``throat'' states would have $E=\tq(\tq+i\tQ)-c_C/12$, which means
that when they exit from the Higgs branch ``throat'' they would have
an energy of
\eqn\energy{E=\tq(\tq+i\tQ)+(c_H-c_C)/12} 
over the Higgs branch vacuum (the ``ground state'' energy of region (4)). 
The mass
gap in the Higgs branch ``throat'' in this sector should thus be
\eqn\massgap{{\tQ^2\over 4}+{{c_H-c_C}\over {12}} = 
{1\over {2N_f}} + {{(6N_f-6)-6}\over {12}} = {Q^2 \over 4},} 
which is the naive mass gap of the Higgs branch. Thus, in this sector
we do not need any vacuum energy to match the Coulomb branch
``throat'' to the Higgs branch ``throat''; the vacuum energy we found
disappears when we do the appropriate shift relating the R-R and NS-NS
sectors. Therefore, our results are in perfect agreement with those of
\seiwit. In the R-R sector of the Higgs branch SCFT
the mass gap (the energy difference between the
``ground state'' and the
continuum) is $1/2N_f$, while in the
NS-NS sector it is $(N_f-1)^2/2N_f$.

\subsec{The ``Throat'' in the Non-Abelian Case}

For $N > 1$, the exact analysis of the ``throat'' becomes more
complicated. However, we still expect integrating out the
hypermultiplets to lead to an effective action for the vector multiplets
which is similar to the action we got in the Abelian
case; gauge invariance and dimensional analysis severely limit the
possible terms that can appear.

The term which we would like to focus on in the non-Abelian case
is a ``commutator'' term, i.e., a potential term which is non-zero when 
the $V$ matrices do not commute. We will not discuss this term
in general, but only in the vicinity of the point in configuration space
in which the $V$ matrices are all proportional to the identity matrix
(this case will be the most interesting for the DLCQ application 
discussed below).
Note that in general
this potential term suggests that the dominant configurations in
the IR are such that the $V$ matrices commute (and can be taken to 
be diagonal). The  
computation of the potential and the effective theory
under such more general circumstances is also 
straightforward, but we will not discuss it here.

Let us expand the matrix $V$ as
\eqn\vexp{V=V_0 \biggl(I + \sqrt{2\over N_f} \delta V \biggr),}
where $I$ is the $N\times N$ identity matrix and we normalized $\delta V$
so it will have a canonical kinetic term in the ``throat''.
Now, we have a double expansion 
in $\delta V$ and in $d(\delta V)/V_0$, 
and the leading potential term in this expansion 
is of the form
\eqn\potcomm{N_f {\tr([V,V]^2) \over {2V_0^2}} = {2\over N_f}
 V_0^2 \tr([\delta V,\delta V]^2).}
This term is related by SUSY to the kinetic term $N_f \tr(dV)^2 / 2 V_0^2$
which we can obtain as in the Abelian case, but we can also compute
it directly.
The mass
squared matrix for the bosonic elements of the hypermultiplets is
a matrix of the form $V^2$, while for the fermionic elements we find
matrices of the form $V^2+i[V,V]$ (with appropriate $SO(4)$
indices). The one-loop computation of the vacuum energy measures
the difference in energy between the ground states of the bosons
and fermions in the hypermultiplet, and we find exactly
the potential \potcomm.

If we have only fundamental hypermultiplets then 
generally, for $N > 1$, a moduli space approximation is
not expected to be valid using the vector multiplet variables\foot{
The moduli space approximation would include only the diagonal
elements of the vector multiplet matrices, integrating out the
off-diagonal elements.}, since
integrating out the off-diagonal vector multiplets gives rise to a
negative metric whenever two eigenvalues approach each other
\diaco. As long as we keep the full vector multiplet matrices
(as above) this problem does not arise. However, we do not fully 
understand the IR dynamics of these vector multiplets. We expect a moduli space
approximation to be valid only when all the eigenvalues of $V$ are
far separated from each other\foot{This is not to be confused with the
situation relevant for DLCQ, which we will discuss momentarily. 
We have so far been discussing the system with 
only fundamental hypermultiplets.
In the DLCQ application there is
an additional adjoint hypermultiplet.}.

\subsec{The ADHM Sigma Model}

For applications described later in the paper to ``little string
theories'' it will be useful to analyze also theories with an adjoint
hypermultiplet, in addition to the $N_f$ multiplets in the fundamental
representation of $U(N)$.  In this case the Higgs branch SCFT is, by the
ADHM construction, the same as 
a sigma model on the moduli space of $N$ instantons
in $SU(N_f)$ (on $\IR^4$).
Note that in this case the problem with the moduli
space of the $V$ matrices
described in the previous paragraph does not arise, since the
positive contribution of the
adjoint hypermultiplet to the moduli space metric cancels
the negative contribution of the vector multiplet. The action for the adjoint
hypermultiplet, which we will denote by $X$, is of the same form as
\higgs\ :
\eqn\higgsadjoint{
{\cal L}_H^{(adj)} = \int d^2x \tr \{ 
(\cD_\mu X)^2 + ([X,V])^2 + \bar{\psi}_X \gamma^\mu \cD_\mu \psi_X +
\bar{\psi}_X [V, \psi_X] + \bar{\psi}_V [X, \psi_X] + D [X, X] \}, }
and the full action of the Higgs branch SCFT
will be the sum of \higgs\ and \higgsadjoint.
Now, we expect to get an effective ``throat'' theory involving both
the vector multiplet and the adjoint hypermultiplet fields, since the
adjoint hypermultiplet does not become massive for large
$V$. 

To derive the effective potential for $X$ in the ``throat'' region 
we note that the
vector multiplet kinetic term which we obtain by integrating out the
hypermultiplets includes (using supersymmetry) a term of the form $N_f \tr(D^2)
/ 2 V_0^2$. Now, integrating out $D$ leads (using \higgsadjoint) to a
potential of the form
\eqn\hyprtrm{{2\over N_f} \tr([X,X]^2) V_0^2.} 
Combining our results, we find that the bosonic potential of 
the adjoint fields in
configurations close to $V=V_0 I$
is of the form
\eqn\bospot{{2\over N_f} V_0^2 \left[ \tr([\delta V,\delta V]^2)
+\tr([\delta V,X]^2)+\tr([X,X]^2) \right].}

\medskip
\noindent{\it The IR Limit of the Effective Action}
\medskip

In the case of the ADHM sigma model, we can proceed further and 
discuss the low-energy limit of the effective action for large $N$. 
In this case, if we discuss the theory on a cylinder,
the lowest energy configurations are given
by ``long string'' configurations, which 
can carry energies proportional to $1/N$ in the
large $N$ limit. Since our ``throat'' theory is a $U(N)$ gauge theory,
with 8 scalar adjoint matrices (4 in $V$ and 4 in $X$), we can
construct such ``long string'' configurations \refs{\motl,\bansei,\dvv}
just like in the DLCQ of
type IIA string theory \refs{\bfss,\taylor,\suss,\grt}, by looking at
configurations which change by a non-trivial $U(N)$ gauge
transformation when going around the circle. 
This relies on the potential term \bospot, 
which forces the low-energy configurations to sit on the moduli space
where all 8 of the matrices commute and can be simultaneously
diagonalized (when we fix $V_0$ and look at configurations with energies
of order $1/N$).

We expect that the
low-energy effective theory on these ``long strings'' will be exactly
the theory we found in the Higgs branch ``throat'' for $N=1$. 
Since the ``long strings'' effectively live on a circle which is
larger by a factor of $N$, 
the mass gap for such configurations is lower by a factor of $N$
than what we found in the $N=1$ case\foot{Recall that we measure energy
in units of the inverse size of the circle.}.
We can also compute the leading correction to this
effective action, which corresponds to the 
 interaction by which ``long strings''
split and join. The analysis is a slight variant of the same analysis 
in the context of the DLCQ of the IIA string
\refs{\motl,\bansei,\dvv}, and the main idea is still that 
this interaction is given by a twist operator \dvv\ whose coefficient is
determined by dimensional analysis.

In the case of the DLCQ of type IIA string theory, the UV is given
by the $1+1$ dimensional $\cn=(8,8)$ $U(N)$ gauge theory. The
theory in that case includes interactions of the form
$g_{YM}^2
\tr([X,X]^2)$, where the 8 $X$ matrices have canonical kinetic
terms. The low-energy theory is an $(\IR^8)^N/S_N$ orbifold, in which
the string coupling may be described as a twist operator. The coefficient
of this operator can be determined by dimensional analysis; since
the operator has dimension 3, its coefficient must scale as
$1/g_{YM}$.

In our case, we claim that the effective theory for ``long strings''
propagating far along the ``throat'' region is similar, but with
$2V_0^2/N_f$ replacing $g_{YM}^2$.  This follows from the effective
potential \bospot\ for configurations where $V$ is close to $V_0 I$.
We conclude that again the
correction to the low-energy theory, expanded around the configuration
$V \simeq V_0 I$ (which breaks conformal invariance), will be governed
by a twist operator in the low-energy orbifold theory, but that this
time its coefficient will be proportional to $\sqrt{N_f/2} / V_0$. 
Note that for small $\delta V$ we can ignore the fact that the $V$'s
really live on $\IR \times S^3$ and not on $\IR^4$.
 
\subsec{Blowing Up the Singularities}

In the previous subsections we saw how a careful analysis of the behavior of the
Higgs branch SCFT near the singularities of the moduli space leads to a
simple description involving a continuous spectrum localized near the
singularities. Another way to analyze the theory near the
singularities, which was used for the $0+1$ dimensional theory in
\abs, is to blow up the singularities of the Higgs branch by adding
Fayet-Iliopoulos (FI) terms of the form $\int d^2x \zeta \tr(D)$ to
the Lagrangian (of course, this is only possible for $U(N)$ gauge
theories, while we expect most of our previous discussion 
to be more general). The
FI term $\zeta$, like $D$, is a triplet of $SU(2)_R$, and it is an
exactly marginal deformation of the Higgs branch SCFT. Adding
such a term lifts the origin of the Higgs branch ($H=0$), and schematically
the minimal allowed value for $H$ now becomes $H \sim
\sqrt{\zeta}$. 
The Coulomb branch is also lifted, so we expect the ``throat'' in the
Higgs branch, which
originally connected the two branches, to no longer be infinite. For
large $N$ this effect was discussed in \seiwit, here we will see how
it works for any value of $N$.

The effect of $\zeta$ on the effective ``throat'' theory is as
follows. As described above, when we integrate out the hypermultiplets
we get a one-loop term proportional to $\tr(D^2) / V_0^2$. Now, if we
integrate out $D$, we find a potential of the form $\zeta^2 V_0^2$.
This potential prevents states from going to large values
of $V$, as expected. In fact, the leading potential which is generated is
precisely the potential appearing in the $\cn=4$ super-Liouville
theory \kpr, as found also in \seiwit. This potential, which preserves
$\cn=4$ superconformal symmetry, can be written using the ``throat''
variables (for $N_f \geq 3$) in the form 
$\zeta^2 \int d^2x d\theta d{\overline\theta}
{\cal O}_{1/2}$, where ${\cal O}_j = e^{j\phi \sqrt{2/N_f}} V_{j,j}$ and
$V_{j,j}$ are the primaries of the bosonic $SU(2)$ level $(N_f-2)$ 
WZW model. It is easy to check that this deformation is marginal, and
it includes
in particular a term which is exponential in $\phi$ and serves as a
``Liouville wall''. 

A similar behavior is expected when we turn on a theta term of the
form $\int d^2x \theta \tr(F_{01})$, which is related by supersymmetry
to the FI terms, but we will not analyze this in detail here. For $U(1)$
with
$N_f=2$ and $\theta=\pi$ we expect to reproduce the free $\IR^4/\IZ_2$
orbifold theory \paul, but we will not attempt to show this here.

\newsec{The DLCQ of $\cn=(2,0)$ ``Little String Theories''}

The rest of this paper will involve applying our results concerning
the Higgs and Coulomb branch SCFTs to the DLCQ of ``little string
theories'' (LSTs) \refs{\natistr,\brs,\olddvv}. 
The original definition of LSTs
with 16 supercharges, as $g_s
\to 0$ limits of NS5-branes or of ALE singularities, is not very useful
for making computations in these theories. However, we now have two
more explicit descriptions which allow computations, one which is a
DLCQ description at finite momentum on a compact light-like circle
\refs{\oens,\edhiggs,\sethi,\ganset}, 
and the other which is a holographic dual of the LSTs \abks\ (see also
\refs{\imsy,\bst}, and see
\refs{\diagom,\kll,\gkpelc,\gk} 
for similar holographic duals to LSTs with less supersymmetry).  The
latter description makes several predictions about LSTs, and we will
attempt here to verify one of these predictions in the DLCQ
description of the LSTs, using what we learned above about the
behavior of the corresponding $1+1$ dimensional field theories.  In
this section we will analyze the DLCQ of $\cn=(2,0)$ supersymmetric
LSTs, and in the next section that of $\cn=(1,1)$ LSTs.

\subsec{Derivation of the DLCQ of $(2,0)$ LSTs}

There are two simple derivations of the DLCQ of the $A_{k-1}$ $(2,0)$
LST \refs{\oens,\edhiggs} (we will not discuss the $D$ or $E$ cases
here, since they seem to present no new issues). The first uses the
definition of the LST as the $g_s
\to 0$ limit of the theory on $k$ NS5-branes in type IIA string
theory. The DLCQ of type IIA string theory (with momentum $P_-=N/R$)
is given by a $1+1$ dimensional $\cn=(8,8)$ $U(N)$ gauge theory on a
circle. The radius of the circle and gauge coupling are given by (up
to numerical constants)
\eqn\symparam{\Sigma = {1 \over R M_s^2}, \ \ \ \ g_{YM}={RM_s^2\over g_s}.}
Longitudinal NS5-branes are described by adding
hypermultiplets in the fundamental representation \berdoug, and the
limit $g_s
\to 0$ corresponds to the limit $g_{YM} \to \infty$, or the IR limit
of the gauge theory. In this limit, the $\cn=(4,4)$ $U(N)$ gauge
theory with an adjoint hypermultiplet and $k$ fundamental
hypermultiplets flows to two different SCFTs, one describing the
Coulomb branch and the other describing the Higgs branch; it is
natural to identify the SCFT of the Higgs branch with the DLCQ of the
decoupled theory on the NS5-branes.

We can derive the same result also directly in the LST, using
Seiberg's description of the DLCQ as equivalent to a compactification
on a space-like circle of radius $R_s \to 0$ with appropriate boosts
to keep the energy finite \matder. In this description we are 
led to discuss the
$A_{k-1}$ $\cn=(2,0)$ ``little string theory'' on a small space-like
circle with $N$ units of momentum on the circle; by T-duality this
is equivalent to the $A_{k-1}$ $\cn=(1,1)$ ``little string theory'' on
a large space-like circle with $N$ strings wrapped on the circle. In
the DLCQ limit we are interested in very low energies in this
configuration; then, we can identify the strings as instantons in the
low-energy $U(k)$ gauge theory, whose low-energy modes are described
by motion on the instanton moduli space, and so we find that the DLCQ
is the $1+1$ dimensional sigma model on the moduli space of $N$ $U(k)$
instantons. Using the ADHM construction, this is equivalent to the
previous result.

As in the $0+1$ dimensional case \abs, we can identify in the DLCQ
description all the space-time symmetries which commute with
$P_-=N/R$. We will not discuss this in detail here since most
identifications are similar to those described in \abs. Let us mention
just how the $SU(2)$ symmetries match between the two theories. The
$SU(2)_r\times SU(2)_l$ symmetry, which is the R-symmetry of the
$\cn=(4,4)$ superconformal algebra, is identified with the
$SO(4)_R$ global symmetry of the six dimensional LST (which originates from 
rotations transverse to the branes). The $SU(2)_R$ symmetry
and an additional $SU(2)_X$ symmetry which is the flavor symmetry of
the adjoint hypermultiplet are identified with the $SO(4)$ symmetry
rotating the four spatial dimensions transverse to the compact
light-like direction.  In the $1+1$ dimensional case there is a much
larger algebra than that which is required by the DLCQ, including
full super-Virasoro and $SU(2)_r\times SU(2)_l$ super-Kac-Moody
algebras. 
The space-time role of the additional generators
is unclear.

\subsec{The ``Throat'' States of the $(2,0)$ LSTs}

\medskip
\noindent{\it The ``Throat'' States in the Holographic Description}
\medskip

The holographic description of the $A_{k-1}$ $(2,0)$
LSTs \abks, which is given by the near-horizon limit of $k$ type IIA
NS5-branes, includes a ``throat''
region given by a linear dilaton background of type IIA string
theory \chs. The full background interpolates between an $AdS_7\times S^4$
background of M theory and the ``throat'' region. The string metric in the 
``throat'' region of the holographic description of the $A_{k-1}$ LST is 
\eqn\throatmetric{ds^2 = dx_{\IR^6}^2 + d\phi^2 + {k\over 2} d\Omega_3^2.}
There is also a 3-form field $H = -kd\Omega$, and the string coupling
behaves as $g_s \propto e^{-\phi\sqrt{2/k}}$, so that the string
world-sheet theory in this region is the sum of a free scalar with
background charge $\tQ = -\sqrt{2/k}$, four free fermions and a bosonic
level $(k-2)$ $SU(2)$ WZW model (and also six more free scalars and fermions
corresponding
to the space-time coordinates of the LST).
The ``throat'' contains a continuum (from the
six dimensional point of view) of states corresponding to
particles (supergravity particles or more general string states) 
with some momentum (incoming,
outgoing or some combination) in the ``throat'' ($\phi$) direction. These
states are delta-function normalizable in the ``throat'', and may
presumably be extended to the strong coupling region in such a way
that they are still delta-function normalizable in the full theory. Thus, we
can take linear combinations of these states that are
normalizable. Since these states can be given any longitudinal
momentum we want (subject to the mass-shell constraint) we should be
able to find these states also in the DLCQ description.

The spectrum of chiral operators in the ``throat'' region includes one
chiral multiplet for every primary field of the corresponding bosonic
$SU(2)$ WZW theory\foot{For the $G$ LST, where $G$ is a group of ADE type,
we have the corresponding $SU(2)$ modular invariant.}. The lowest
multiplet is the supergraviton multiplet. For this multiplet
the on-shell condition is of the form 
\eqn\onshellcond{E^2 - p^2 - \tq(\tq+i\tQ) = 0,}
where $E$ is the energy,
$p$ (a 5-vector) is the spatial momentum and $\tq$ is the
momentum in the $\phi$ direction (i.e. the vertex operator is $e^{i
\tq \phi}$), all dimensionless and measured in string units. The S-matrix
for these states was discussed in \minwalla.

In the DLCQ of this theory, 
we should find such states with longitudinal momentum
$p_-=N/R$ and with any $\tq$ and transverse momentum $\vec{p}$. The
DLCQ Hamiltonian $p_+$ for these states should be of the form 
\eqn\dlcqham{p_+ =
(\vec{p}^2 + \tq(\tq+i\tQ)) / p_- = R (\vec{p}^2 +
\tq(\tq+i\tQ))/N.}
The minimal value of this Hamiltonian, which occurs
for $\vec{p}=0$ and $\tq=-i\tQ/2$, is given by
$R\tQ^2/4N = R/2Nk$. Similar
equations (with a larger mass gap) arise for the other chiral
multiplets.

In addition to the spectrum, we also know how the string interactions
are supposed to behave like in the ``throat''. The linear dilaton
causes the string coupling to behave as $g_s \propto e^{\tQ\phi}$,
and we should be able to reproduce this behavior in the DLCQ.

\medskip
\noindent{\it The ``Throat'' States in the DLCQ Description}
\medskip

Matching the spectrum and string interactions in the ``throat'' region
between the
DLCQ and the holographic description is straightforward
in view of the field theoretic analysis in sections 2 and 3. 
As an initial step it is interesting to compare the Higgs branch scaling
of $V$ with the near-horizon scaling of the coordinates \berver.
In DLCQ the Lagrangian for the coordinates $r^i$ transverse to the brane
is naturally of the form 
\eqn\dlcqlag{{\cal L}_{DLCQ} \sim {1\over R\Sigma} \int d^2x (\partial_\mu
 r^i)^2,}
corresponding to a low-energy quantum-mechanical Hamiltonian proportional
to $R$.
The relation to the SYM Lagrangian \gaugetheory\ variables is thus given by
\eqn\reltrvs{V^i={g_{YM}r^i \over \sqrt{R\Sigma}}=
R \biggl({M_s^3\over g_s} r^i\biggr).}
The quantity in parenthesis is precisely what we keep fixed in the 
near-horizon limit \refs{\imsy,\abks}
of NS 5-branes\foot{It is the tension of the strings that
originate from stretched D2-branes.}. Hence, keeping the dimension one $V$ fixed
is equivalent to taking the near-horizon limit.

We begin our discussion of the ``throat'' states in the DLCQ description
with the simplest case of the $A_{k-1}$ LST with
$N=1$. The DLCQ is then simply the Higgs branch SCFT of the $U(1)$
gauge theory with $N_f=k$ hypermultiplets, discussed earlier.
In addition there is also an adjoint hypermultiplet,
which in the $U(1)$ case is just a free hypermultiplet, giving rise to
states of momentum $\vec{p}$ with energy $\vec{p}^2$. We found above
that this theory has a region in moduli space which looks like a
``throat'', involving a supersymmetric level $k$ $SU(2)$ WZW model and
a scalar field with background charge $Q = (k-1) \sqrt{2/k}$. We also
found a zero-point energy in these variables, so that the Hamiltonian
\higgsham\
is of the form $H_{DLCQ} = (1 / \Sigma) (\vec{p}^2 + q(q+iQ) + 1 -
k/2)$, where we reintroduced the correct units for the Hamiltonian by
using the radius of the circle in the SCFT, which is $\Sigma = 1 / R
M_s^2$. As described at the end of section 3.1, we can define $\tq = q +
i(Q-\tQ)/2$ (where $\tQ=-\sqrt{2/k}$ as above), and
rewrite this Hamiltonian as $H_{DLCQ} = R M_s^2 (\vec{p}^2 + \tq (\tq
+ i\tQ))$. This is exactly the same equation we found above for $p_+ =
H_{DLCQ}$ (for $N=1$). Thus, we find the same continuum of ``supergraviton''
states in both descriptions of the LSTs. In section 3 we described the change
of variables from $q$ to $\tq$ in the context of relating the Higgs
branch and Coulomb branch ``throats'' of the same theory; here we see
that the same change of variables relates the DLCQ Higgs branch
variables and the space-time variables, even though there is no direct
relation (in the DLCQ context)
between the space-time ``throat'' and the Coulomb branch
after we take the decoupling limit. We can easily generalize this
analysis to the other chiral states of the LSTs. In both descriptions
they are in a one-to-one correspondence with the primaries of the
(same) bosonic $SU(2)$ WZW model, and it is easy to check that we find
exactly the same states in both cases.

For higher $N$ the situation is similar, for states localized near
some position in the ``throat'' which are the configurations we discussed
in section 3.2. As in the DLCQ of
critical string theories, we need to look at ``long string''
configurations that can carry energies proportional to $1/N$ in the
large $N$ limit. We have discussed in section 3 how to go to the ``long
strings'' for large $N$ and what is the coefficient of
the twist operator in the effective theory. 
Translating the results there to the language of the current section leads
to exactly the correct Hamiltonian for the ``throat'' states, and to a
string coupling in space-time of the form
\eqn\stringcoup{g_s \propto 1/V_0 \simeq e^{-\sqrt{2/k}
\phi}.}
This is exactly the same string coupling found in the
holographic description of the ``little string theories'' \abks\ (note
the factor of $\sqrt{2}$ difference in the normalization of $\phi$
between our paper and \abks). Thus, we see that both the low-energy
states (in the ``throat'')
and their interactions are correctly reproduced in the
DLCQ.

\subsec{Blowing Up the Singularities in the DLCQ}

In section 3.3 we discussed blowing up the singularities in the Higgs branch
by turning on FI terms (or theta angles) in the gauge theories, and we
showed that this turns on a ``Liouville wall'' in the ``throat'' which prevents
states from propagating deep into the ``throat'' (this description is useful
for small FI terms, when the
``throat'' approximation is still valid near the ``wall''). In the DLCQ
context we can give this blow-up a space-time interpretation. Following
the derivation of the DLCQ, using the definition of the LST as the $g_s
\to 0$ limit of NS5-branes, we can show that the FI terms correspond 
in this definition to turning on 3-form RR fields parallel to the NS5-branes
(which are non-trivial, despite having zero field strength,
due to their interaction with the 3-form fields on
the NS5-branes). This is a generalization of the analogous deformation of
the $(2,0)$ six dimensional SCFT discussed in \refs{\abs,\michanoncom}
(and more precisely
in \newseiwit), and of the deformations used for constructing 
non-commutative Yang-Mills theories. 

Presumably, the decoupling limit of the NS5-branes in the presence of this
RR field leads to some analog of a non-commutative ``little string
theory''. Since we identified the ``throat'' of the Higgs branch with the
linear dilaton region in the holographic dual of the LSTs, we see that
turning on the deformation turns on a wall that prevents states from going
into the weak coupling region in space-time; 
thus, we expect this deformation in the
holographic description to leave the $AdS_7\times S^4$ region and the
strong coupling part of the linear dilaton region, but to lift the weak
coupling region (this is opposite from the standard ``Liouville wall''). 
This should be visible also in the appropriate
near-horizon limit of the NS5-brane with RR fields (generalizing the
results of \refs{\akisanny,\juanrusso}).
This is opposite to the deformations recently discussed
in \gkpelc; they discussed deformations which change the IR behavior of
the LSTs, while the deformation discussed here does not change the
IR behavior but seems to significantly change the UV behavior. It would
be interesting to study these theories further.

\newsec{The DLCQ of $\cn=(1,1)$ ``Little String Theories''}

\subsec{Derivation of the DLCQ of $(1,1)$ LSTs}

To derive the DLCQ of the $A_{k-1}$ $(1,1)$ LSTs, it is simplest to
start from their definition as the $g_s \to 0$ limit of the theory on
an $A_{k-1}$ singularity in type IIA string theory. In \refs{\sethi,\ganset}
it was shown that the DLCQ
description of this is given by the low-energy theory of D-strings
near a similar singularity in type IIB string theory\foot{To
derive this most directly, we start with this theory on a small space-like
circle with $N$ units of momentum (and with finite $g_s$); it is
natural to T-dualize this to type IIB string theory (with the same
singularity) on a large circle with $N$ fundamental strings, but now
we get a large coupling for the type IIB string theory so it is
natural to S-dualize, and we end up with $N$ D-strings near an
$A_{k-1}$ singularity at weak coupling. The $g_s \to 0$ limit is again the
low-energy ($g_{YM} \to \infty$) limit.}. Unfortunately, there is no
simple description of this theory. However, if we deform the
singularity to the $\IR^4/\IZ_k$ orbifold, then the
theory on the D-strings
is \dougmoore\ the $\cn=(4,4)$ $U(N)^k$ SQCD theory
with bi-fundamental hypermultiplets for adjacent group factors (when
we arrange the gauge groups on a circle). At the orbifold point the
gauge couplings of all $U(N)$ factors are equal and they
are inversely proportional to the original type IIA string
coupling. Thus, the $g_s \to 0$ limit again corresponds to taking the
IR limit of this theory, which flows to two decoupled SCFTs. In this
case it is the Coulomb branch theory which we identify with the DLCQ
of the LST, since the Higgs branch describes the motion of the strings
away from the singularity.

One still needs to discuss the effects of the deformation from 
the $A_{k-1}$ singularity to the non-singular $\IR^4/\IZ_k$
orbifold. This deformation corresponds to turning on specific 
non-zero $B$ fields on the various vanishing 2-cycles of the singularity. 
If we follow the chain of dualities, we find that in the original 
description turning on these $B$
fields corresponds to turning on a longitudinal Wilson line in the
low-energy $U(k)$ gauge theory\foot{The $B$ fields correspond to the
longitudinal components of the RR vector fields in the twisted sector
of the $\IR^4/\IZ_k$ orbifold, which are the Cartan subalgebra of the
low-energy
$U(k)$ theory.}. More precisely, the orbifold point corresponds to 
a Wilson line of the form
\eqn\wilsonline{RA_- = \pmatrix{0 & 0 & 0 & \cdots & 0 \cr
                              0 & {1 \over k} & 0 & \cdots & 0 \cr
                              0 & 0 & {2 \over k} & \cdots
                              & 0 \cr
                              \vdots & \vdots & \vdots & \ddots &
                              \vdots \cr
                              0 & 0 & 0 & \cdots & {{k-1}
                              \over k} \cr}.}
This Wilson line has the effect of shifting the longitudinal momentum
of charged particles. The fields in the Cartan subalgebra
(corresponding to diagonal matrix elements)
still have integer longitudinal momentum, but the other fields now have
a fractional momentum; the space-time fields (e.g. W bosons) coming
from the $(i,j)$ element of a $U(k)$ matrix have $RP_- = (i-j)/k$ 
(mod $1$).

Changing the value of this Wilson line corresponds to changing the $B$
fields in the $A_{k-1}$ singularity used to define the DLCQ theory. In
the field theory of the D-strings, this corresponds to changing the
ratios between the Yang-Mills couplings of the $k$ $U(N)$ gauge
groups. In particular, turning off the Wilson line (or some components
of it, leading to an unbroken low-energy non-Abelian gauge group in
space-time) corresponds to
making ratios of gauge couplings infinite. At first sight, since to
describe the LST we are taking the gauge couplings to infinity, these
ratios of gauge couplings seem to be unimportant; however, we will see
below that they show up in the moduli space metric.

The identification of the 
$SU(2)$ symmetries between the DLCQ field theory and the space-time LST is the
following. 
The $SU(2)_r\times SU(2)_l$ group in the two dimensional gauge theory is now
identified with the longitudinal $SO(4)$ of the six dimensional theory
in the light-cone frame, while $SU(2)_R$ is
identified with a diagonal subgroup of the global $SO(4)_R$ symmetry group
of the LST.
As discussed above,
we expect that in the Coulomb branch SCFT the $SU(2)_R$ symmetry will
be enhanced to an $SU(2)\times SU(2) \simeq SO(4)_R$ Kac-Moody algebra, 
although we cannot
exhibit this symmetry directly in the gauge theory.

\subsec{The Low Energy Theory}

\medskip

\noindent{\it The Metric on the Moduli Space}

\medskip

The moduli space of the Coulomb branch is given by the scalars $V$ in
the $U(N)^k$ vector multiplets; the four scalars of each $U(N)$ factor
are commuting on the moduli space, so they can be simultaneously
diagonalized. We can thus parametrize the moduli space by the vectors
$\vec{V}_i^{(j)}$, where $V$ is a 4-vector, $i=1,\cdots,k$ labels the
different $U(N)$ groups, and $j=1,\cdots,N$ labels the eigenvalues of
the corresponding matrices (in some order). The moduli space metric
gets contributions only from tree level and 1-loop. Thus, following
\diaco, it is easy to compute the exact metric,
which is
\eqn\metric{ds^2 = {1\over g_{YM}^2} \sum_{i=1}^k \sum_{j=1}^N (d{\vec
V}_i^{(j)})^2 + \sum_{i=1}^k \sum_{j,l=1}^N 
{{d({\vec V}_i^{(j)}-{\vec V}_{i+1}^{(l)})^2} \over
|{\vec V}_i^{(j)}-{\vec V}_{i+1}^{(l)}|^2} -
\sum_{i=1}^k \sum_{j=1}^{N-1} \sum_{l=j+1}^N 
{{2d({\vec V}_i^{(j)}-{\vec V}_i^{(l)})^2} \over
|{\vec V}_i^{(j)}-{\vec V}_i^{(l)}|^2},}
where $(i+1)$ is taken modulo $k$. 

In the DLCQ of the LST we are interested in the limit $g_{YM} \to
\infty$, so naively the tree level term drops out; however, as discussed in
section 2, we
should also take the $V$'s to infinity such that $\Phi \sim V/g_{YM}$
remains constant in the IR limit. This can be seen in various
ways. We justified this in section 2 by requiring zero dimension for
the moduli space variables, but in the DLCQ context we can also see
this by noting that the space-time distance between states is
proportional to $\Phi$ and not to $V$.
It will be convenient to normalize $\Phi
\equiv \sqrt{k} V / g_{YM}$ (because $g_{YM}$ in this theory is
actually $\sqrt{k}$ times what it was for the D-strings without the
orbifold), and then we find that the actual metric which is relevant
for the DLCQ of the LST is of the form
\eqn\metric{ds^2 = {1\over k} \sum_{i=1}^k \sum_{j=1}^N (d{\vec
\Phi}_i^{(j)})^2 + \sum_{i=1}^k \sum_{j,l=1}^N 
{{d({\vec \Phi}_i^{(j)}-{\vec \Phi}_{i+1}^{(l)})^2} \over
|{\vec \Phi}_i^{(j)}-{\vec \Phi}_{i+1}^{(l)}|^2} -
\sum_{i=1}^k \sum_{j=1}^{N-1} \sum_{l=j+1}^N 
{{2d({\vec \Phi}_i^{(j)}-{\vec \Phi}_i^{(l)})^2} \over |{\vec
\Phi}_i^{(j)}-{\vec \Phi}_i^{(l)}|^2}.} 
As mentioned above, if we change the longitudinal Wilson line in
space-time, we need to change the ratios of Yang-Mills couplings in
the DLCQ description; we see that in the moduli space metric this
corresponds to having different coefficients for the $d{\vec \Phi}^2$
terms corresponding to different gauge groups (the coefficients in the
normalization we chose are exactly the values of the $B$-fields in a
particular basis for the 2-cycles).  In particular, if we try to turn
off some of the Wilson line components, some of the $d{\vec
\Phi}^2$ terms vanish, leaving only the 1-loop contributions (and changing
the asymptotic form of the moduli space).

\medskip

\noindent{\it The Singularities}

\medskip

In the presence of the Wilson line, the metric is flat when all the
eigenvalues are far from each other, but develops various singularities
when some of the eigenvalues come close together. When a vector ${\vec
\Phi}_i^{(j)}$ approaches a vector ${\vec \Phi}_{i+1}^{(l)}$ the
theory looks like a $U(1)$ theory with one charged
hypermultiplet. Naively this theory has a ``throat''-like singularity, but
in fact the moduli space description in the ``throat'' breaks down \diaco,
and there is no good description of this singularity. We do not expect
to have a
``throat'' emerging from this
sort of singularity; at least, there is no other branch emanating from
such a point (before taking the IR limit) which would indicate such a
``throat''. It is believed that this singularity is smoothed
out \refs{\diaco,\natipri}.

A hint to the way by which this singularity is smoothed out is given
by the fact that at low energies (i.e, restricting to the quantum 
mechanics of the zero-modes\foot{We assume that supersymmetry protects
the zero-mode dynamics from being renormalized by integrating out the non-zero
momentum modes.})
it is known that there is a bound state
living at this singularity (the D0-D4 bound state). This suggests 
that there is also a ground state of the $1+1$ dimensional theory
which is localized near the 
singularity (this was referred to as a ``quantum Higgs branch'' in \edhiggs). 
We will later motivate this expectation using
DLCQ considerations.

Additional singularities arise when several vectors ${\vec
\Phi}_{i+n}^{(l_n)}, n=0,\cdots,j$, $(j < k-1)$
approach each other. There is no good
description of these singularities, but by analogy with the
previous case we would expect the moduli space description to
break down, and the existence of a bound state at low energies
(again, this expectation will be supported by the DLCQ analysis below).

Another type of singularity occurs when a vector ${\vec
\Phi}_i^{(j)}$ approaches a different vector ${\vec
\Phi}_i^{(l)}$. In this case the effective theory is like the pure
$SU(2)$ theory, for which again the moduli space description breaks
down since the metric becomes negative \diaco. Also in this case it is believed
that there is no ``throat'' emanating from such a singularity, since there
is no other branch coming out of it prior to taking the low-energy limit. 
Presumably, this singularity is also smoothed out in appropriate variables.

A more serious singularity
arises when $k$ vectors coming from different gauge groups approach
each other; in this case we expect a real ``throat'' to develop in the
moduli space (as is evident in the metric), since such a singularity
connects (before taking the IR limit) to the Higgs branch of the
theory, in which the hypermultiplets acquire VEVs. 
We will discuss this ``throat'' in section 5.3.

\medskip

\noindent{\it DLCQ Applications}

\medskip

For now, let us return to the question of the low-energy states. We
will analyze them in detail for $N=1$; the analysis of other cases is
similar for configurations where the different eigenvalues of each
matrix are far from each other (giving $N$ copies of the $N=1$ case),
and is not clear for other configurations. 

The simplest configuration
involves slow motion on the asymptotic region of the moduli space,
where all $k$ of the ${\vec \Phi}$'s are far from each other and the
metric is flat. Restricting to the zero modes on the circle, 
we identify this configuration with a space-time
configuration involving $k$ W-bosons\foot{We use the term W-boson to
denote any element of the space-time vector multiplet. The precise
element in space-time will be determined by the fermion zero modes.},
coming from the $(1,2),(2,3),\cdots,(k-1,k)$ and $(k,1)$ elements of
the low-energy $U(k)$ matrices.
Each vector ${\vec \Phi}_{i}$ may be identified with
the transverse position of one of these particles, each of which
carries $1/k$ units of longitudinal momentum (such that the total
longitudinal momentum is $P_-=1/R$), and the Hamiltonian is
consistent with this interpretation\foot{A related discussion of
W-boson states appears in \ganset.}. 
If we
change the longitudinal Wilson line, the description remains the same;
the coefficients of the $d{\vec \Phi}^2$ terms in the moduli space
metric now change in the asymptotic region, corresponding to the
different longitudinal momenta now carried by the $k$ particles, and
we still get an exact agreement between the space-time and DLCQ
theories. If we try to turn off a component of the
Wilson line we see that this flat
region of the metric goes off to infinity, as we expect since then one
of the particles we describe has $P_- \to 0$ and we would not expect
to describe it as a simple particle in the DLCQ. From here on we will
discuss only the ``maximal'' Wilson line
\wilsonline.

We still need to account for the rest of the W-bosons in 
space-time. Our proposal is that these are described by bound states
which are the same bound states discussed above in the context of
resolving the singularities in the moduli space.
For example, the W-boson corresponding to
$(1,3)$ elements of the matrices may be viewed as a bound state at
threshold of the $(1,2)$ and $(2,3)$ W-bosons, so we identify it
with a configuration in which two adjacent vectors, say ${\vec
\Phi}_1$ and ${\vec \Phi}_2$, come together and form a bound state in
the region where naively there is a ``throat''. From the space-time
description we see that the effective dynamics of this bound state
should be (when it is far from the other states) just a sigma model on
$\IR^4$, with a metric whose coefficient is twice that of the original
two vectors. This would reproduce the space-time dynamics of a
particle with $RP_-=2/k$, as desired. Similarly, by taking various
combinations of such bound states, we can describe any other
configuration of W-bosons whose total momentum is $RP_-=1$. 

 From this description it is clear that states corresponding to photons
(vectors in the Cartan subalgebra), which have $RP_-=1$, live in the
``throat'' region, where all the eigenvalues come together. 
We will discuss this region in the next subsection.

\subsec{The ``Throat'' States of the $\cn=(1,1)$ LST}

In the $\cn=(1,1)$ case the Coulomb branch metric receives one-loop
corrections that cause it to include semi-infinite ``throats'', so it seems
that it may be possible to identify the ``throat states''
more directly than in the $\cn=(2,0)$ case discussed above. 
However, surprisingly, it actually seems more complicated to
obtain a precise understanding of the ``throat states'' in this case.

There is one case in which we can precisely identify the ``throat
states'', which is the $A_1$ $\cn=(1,1)$ LST with one unit of
longitudinal momentum. 
 From the holographic
description \abks\ we learn that in this case there is (since the
bosonic WZW model in the space-time string theory has level
$k_{SU(2)}=2-2=0$) just one chiral multiplet of states propagating in
the ``throat'', which is the graviton multiplet. The scalar $\phi$
parametrizing the ``throat'' coordinate has a background charge of
$\tQ=-\sqrt{2/k}=-\sqrt{2/2}=-1$. The on-shell condition (or the condition
for the corresponding vertex operator to be of dimension $(1,1)$) for
these states is $E^2 - p^2 - \tq(\tq+i\tQ) = 0$, where $p$ (a 5-vector) is
the spatial momentum and $\tq$ is the momentum in the $\phi$ direction
(i.e. the vertex operator is $e^{i\tq\phi}$), all measured in string
units.

For $N=1$ we can find exactly the same states in the DLCQ.
The DLCQ description of this theory is the IR
limit of the Coulomb branch of the $U(1)\times U(1)$ $\cn=(4,4)$ gauge
theory with 2 bifundamental hypermultiplets, which 
is equivalent to a free $U(1)$ gauge theory (which is just
a sigma model on $\IR^4$) and another $U(1)$ theory with two charged
hypermultiplets. As described in \diaco, when
we integrate out these hypermultiplets the metric develops a ``throat''. In
the ``throat'', the theory of $U(1)$ with $N_f=2$ is a single boson $\phi$
with background charge $\tQ=-\sqrt{2/N_f}=-\sqrt{2/2}=-1$, and (after the
chiral twist of the fermions) a bosonic WZW model at level
$k_{SU(2)}=2-2=0$ (there are also some free fermions). Thus, in the
DLCQ we can construct states (localized in the ``throat'' region) with an
arbitrary transverse 4-momentum ${\vec p}$ (from the free $U(1)$ part)
and an arbitrary $\phi$ momentum $\tq$. Such a state will have a DLCQ
Hamiltonian of the form
\eqn\hamilt{H_{DLCQ} = R [{\vec p}^2 + \tq
(\tq + i\tQ)],} 
so that 
\eqn\massshell{p_+ p_- - {\vec p}^2 = H_{DLCQ} * 1 / R - {\vec p}^2 = \tq (\tq +
i\tQ),} 
which is exactly the same answer we found above from the
holographic description. Thus, in this case we have exactly the right
``throat states'' (the fermion zero modes give these states the
correct multiplicities in space-time). Note that in space-time the
dilaton becomes smaller as we go out in the ``throat'' (away from the
5-branes), while in the DLCQ description it becomes larger as we go 
into the ``throat'' ($\phi$ in the DLCQ is $(-\phi)$ in space-time), 
but there is no obvious relation between the two
dilatons (as discussed above, the string interactions in space-time
are related to a different world-sheet operator).

For higher values of $N$ and/or $k$ things become much more
complicated. Let us start by discussing higher $k$. In this case,
ignoring the free $U(1)$ multiplet, the
$4k-4$ dimensional moduli space has a co-dimension $4k-4$ singularity at
which, as described above, the metric seems to develop a semi-infinite
``throat''. However, for $k > 2$ it is not clear how to go to a simpler
description of the ``throat'' theory as we did for $k=2$. We expect to
still have one coordinate labeling the distance ``down the throat'',
but it is not clear
what would be a good description of the $4k-5$ angular variables in
this ``throat'' (or whether some of them decouple). 
One expects these variables to be equivalent to a level $(k-2)$ bosonic
$SU(2)$ WZW model, corresponding to the ``throat'' theory we have in
space-time (similar to what we found for the $(2,0)$ LST). However, except
for ``answer analysis'' and the fact that this description has the
correct $SU(2)\times SU(2)$ symmetries (which we have in
space-time, as discussed in the next paragraph), there is no good
argument for it. 
Encouraged by the clean picture that arises in the $(2,0)$ case,
we will view the ``throat'' states of the holographic
description as a prediction about the behavior of ``throat'' states in
the Coulomb branch of these gauge theories.

The $SU(2)$ that is part of the $SU(2)\times SU(2)$ symmetry of
the putative $SU(2)$ WZW model
in the effective theory of the ``throat'' is what we called before $SU(2)_R$.
Since this group acts only on fermions in the Coulomb branch, 
the WZW model cannot come directly
from the bosonic moduli space coordinates, but must involve
the fermionic variables. Presumably, again the appropriate ``throat''
variables are some combinations of the fermions.
By now this is not surprising,
since this is how we got the ``throat''
WZW model in the Higgs branch, and similar
``bifermionic coordinates'' were found 
in the DLCQ of $\cn=(2,0)$ SCFTs \berver\ and in instanton computations on
$AdS_5\times S^5$
(see \inst\ and references therein). 

For higher values of $N$ we again expect to find ``long string''
configurations which can be identified with the strings propagating
in the ``throat'' of the holographic description. Since on the
Coulomb branch we have a $U(N)^k$ gauge symmetry it is easy to
construct ``long string'' configurations which vary by a gauge
transformation when going around the circle, but we do not know how to
show that they are described by the correct theory and that they have
the correct interactions (as we showed for
the $\cn=(2,0)$ case above).

\newsec{Summary and Discussion}

In this paper we analyzed the superconformal theories arising as the low-energy
limit of $\cn=(4,4)$ gauge theories. We wrote down an explicit
Lagrangian for the Higgs branch SCFT, and showed how it leads on one
hand to the description of this SCFT as the sigma model on the Higgs
branch moduli space, and on the other hand to ``throat'' states
localized near the singularities of this space. We analyzed these
``throat'' states in detail for the cases of $U(1)$ and $U(N)$ gauge
theories, and matched them in the DLCQ context to ``throat'' states 
of the $(2,0)$ LSTs. It would be interesting to generalize this
analysis to higher order terms and to different gauge groups.
Our analysis of the $U(1)$ theory with two hypermultiplets applies
to the $\IR^4/\IZ_2$ sigma model with zero theta angle. A similar analysis
of $\IR^4/\IZ_k$ singularities requires studying $U(1)^k$ theories with
bifundamental matter, which we leave for further study.
We could not write down an explicit Lagrangian for the Coulomb branch SCFT,
but we wrote down its moduli space metric and used it to analyze some states
in the SCFT in section 5. 

The Coulomb and Higgs branch SCFTs of the same gauge theory both
contain (when they both exist) isomorphic ``throat'' regions, though the
full theories are quite different and have different asymptotic regions. The
``throat'' theory has a large $\cn=4$ algebra, and different small $\cn=4$
subalgebras of this are realized as the superconformal symmetries of the
Higgs and Coulomb branch theories.

It is important to note that the ``throat'' states generally account
for just a small
fraction of the density of states in the Higgs branch SCFT. For example,
in the case of $U(1)$ with $N_f$ hypermultiplets, the density of states
of the full theory is governed by the central charge $c=6(N_f-1)$, while
the density of ``throat'' states is that of a $c=6$ theory (the ``throat''
theory in itself does not obey Cardy's formula for
the asymptotic density of states). This may be related to the fact
that in the ``throat'' we can only see states that are invariant under
the $SU(N_f)$ flavor symmetry of the Higgs branch theory.

In the DLCQ context the density of states in the Higgs branch SCFT (for
the case of $U(N)$ gauge group with an adjoint and $k$ fundamental
hypermultiplets) is
related to the density of states in the $A_{k-1}$ 
LSTs at high energy densities
\ahaban, which in turn is related to the density of states of
CGHS black holes \refs{\cghs,\andyjuan}. 
The relevant states in the Higgs branch SCFT are those whose energy scales
as $1/N$ in the large $N$ limit.
As discussed in \ahaban, in order to reproduce the black hole entropy one
needs the density of these states to behave like that of a conformal theory
with $c=6k$. 
In the sigma model description of the Higgs branch theory it is hard to
see states with an energy scaling as $1/N$; however, we saw in section 3
that in the ``throat'' region we have states whose energy scales like $1/N$,
which are the ``long string'' states. Unfortunately, as discussed above,
the density of these states is much smaller (for $k > 2$) than the full
density of states which we need to reproduce the black hole entropy. 
We predict that appropriate states exist in the full Higgs branch SCFT,
perhaps corresponding to states with non-trivial flavor $SU(k)$
transformations, but we do not know how to exhibit them directly. It would
be interesting to find these states and try to use the DLCQ description to
learn about the high-energy behavior of the LSTs. 

\bigskip

\centerline{\bf Acknowledgements}

We would like to thank M. Aganagic, V. Balasubramanian, T. Banks,
J. Brodie, S. Kachru, A. Kapustin, A. Karch, D. Kutasov, R. Plesser,
A. Rajaraman, M. Rozali, S. Sethi, S. Shenker, E. Silverstein, 
A. Strominger, and especially
N. Seiberg and H. Verlinde for useful discussions.  The work of
O.A. was supported in part by \#DE-FG02-96ER40559, and that of
M. B. by NSF grant 98-02484. We would like to thank the Aspen Center
for Physics for hospitality during the final stages of this work.

\listrefs

\end